\documentclass[aps,prb,preprint, superscriptaddress]{revtex4-2}
\usepackage[]{graphicx}
\usepackage{amsmath}
\usepackage{amsfonts}
\usepackage{amssymb}
\usepackage{natbib}
\usepackage[version=4]{mhchem}
\DeclareMathOperator*{\argmin}{arg\,min}
\begin{document}
\title{Efficient method for magnetic structure exploration based on
first-principles calculations:
application to MnO and hexagonal ferrites \ce{SrFe12O19}}
\author{Taro Fukazawa}
\author{Haruki Okumura}
\author{Tetsuya Fukushima}
\affiliation{Materials DX Research Center, National Institute of Advanced Industrial Science and Technology,
Tsukuba, Ibaraki 305-8568, Japan}
\author{Hisazumi Akai}
\affiliation{Graduate School of Engineering, The University of Osaka, Suita, Osaka 565-0871, Japan}
\author{Takashi Miyake}
\affiliation{CD-FMat, National Institute of Advanced Industrial Science and Technology,
Tsukuba, Ibaraki 305-8568, Japan}
\date{\today}
\begin{abstract}
We propose an approach for exploring magnetic structures 
by using
Liechtenstein's method for exchange couplings from the results
of first-principles calculations.
Our method enables efficient and accurate exploration of stable magnetic structures
by greatly reducing the number of 
first-principles calculations required. We apply our method to the
magnetic structures of MnO and hexagonal ferrite
\ce{SrFe12O19}.
Our method
correctly identifies the ground-state magnetic structure
with a small number of first-principles calculations in these systems.
\end{abstract}
\maketitle
\section{Introduction}
Magnetic materials exhibit a rich variety of physical properties,
including ferromagnetism, antiferromagnetism, and ferrimagnetism.
Understanding the magnetic structure of materials is essential
for the design and development of functional magnetic materials.

\begin{figure}
    \centering
    \includegraphics[width=7cm]{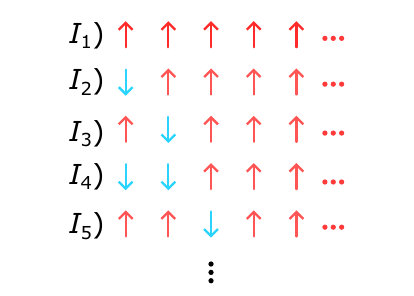}
    \caption{\label{trial_states}
    Schematic example of trial states
    (denoted by $I_1, I_2, \cdots$) in a search for the ground-state magnetic structure.
    }
\end{figure}

First-principles calculations
based on density functional theory (DFT) \cite{Hohenberg64,Kohn65}
are widely used to search for magnetic materials.
One difficulty in treating magnetic systems is the existence of
metastable magnetic structures.
To find the ground state of the system, the total energy
of the trial magnetic states must be calculated and compared.
This typically needs manual handling of the initial magnetic state and 
repeated DFT calculations
for finding the ground state of the system.
The space for the trial states (or the search space) can be vast,
and computational cost is a particular concern
when there is a large degree of freedom in the search
(Fig.~\ref{trial_states}).
To reduce the computational cost of magnetic structure exploration,
several approaches are possible.
For example,
a genetic algorithm has been used to search
for magnetic materials \cite{Ishikawa20,Ishikawa21},
and Bayesian optimization has been used to search for crystal structures.  \cite{Yamashita18}
However,
these methods still
require many first-principles calculations.
In this study, we propose the construction of an efficient 
magnetic model
from a single first-principles calculation
using Liechtenstein's method \cite{Liechtenstein87} for inter-site magnetic
couplings.
We validate the method by 
searching for the most stable structures of 
\ce{MnO} and hexagonal
ferrite \ce{SrFe12O19}, which have complex magnetic structures.

The remainder of this paper is organized as follows.
In the next section, we describe the methodology involved in the study,
including the method for searching for magnetic structures and the
calculational setup for the first-principles calculations.
In Section 3, we present the results for
MnO and hexagonal ferrite \ce{SrFe12O19}, and we discuss
the efficiency of our approach in identifying the ground-state
magnetic structure.
Finally, in the Conclusion section,
we summarize our findings and discuss the potential of our approach
to reduce the computational cost of magnetic structure exploration.

\section{Method}
\subsection{Magnetic structure search}
\label{method_magsearch}
In our magnetic structure search,
we use the classical Heisenberg model specified by $J_{i,j}$,
\begin{equation}
    \mathcal{E}
    =
    \sum_{i,j}
    J_{i,j}\,
    \vec{e}_i
    \cdot
    \vec{e}_j,
    \label{eq:Heisenberg}
\end{equation}
where $\vec{e}_i$ is a unit vector that denotes the direction of
the local magnetic moment.
This equation can be transformed to the ordinary form of the Heisenberg Hamiltonian,
$\mathcal{E} = \sum_{i,j} \tilde{J}_{i,j}\, \vec{S}_i \cdot \vec{S}_j$,
with $\tilde{J}_{i,j} = J_{i,j}/(S_i S_j)$.

We calculate 
the $J_{i,j}$ values by using the method described in the next section.
These $J_{i,j}$ values depend on the final magnetic state
obtained by the first-principles calculation.
Let $\mathcal{S}=(e_1, e_2, ..., e_M)$
denote the final magnetic state with which the $J_{i,j}$ variables are calculated.
Because
spin-collinearity is assumed in the present calculation,
we can describe the final magnetic structure with Ising-like states,
$e_i = {\pm}1$, corresponding to 
the parallel and antiparallel local moment to the $z$-direction
for the $i$th site in the cell.
We distinguish atoms by their Wyckoff positions in
$\mathcal{S}=(e_1, e_2, ..., e_M)$.
In this case, $M$ is the number of crystallographically
inequivalent atoms.
We express the dependency of $J_{i,j}$ on the final magnetic structure
as
$J_{i,j} (\mathcal{S})$.

In statistical atomistic spin simulations,
the Heisenberg model, Eq.~\eqref{eq:Heisenberg}, is assumed to reproduce
the values of the relative energy adequately in DFT among trial magnetic states.
Let $I$ denote a trial magnetic state, the energy function in DFT by $E_{\mathrm DFT}$,
and the energy function from the Heisenberg model by $\mathcal{E}(I,S)$.
The energy from the model also depends on the magnetic state indicated by $\mathcal{S}$.
We can express this assumption mathematically as
\begin{equation}
    E_{\mathrm DFT}(I)
    \simeq
    \mathcal{E}(I,\mathcal{S})
    +
    C(\mathcal{S}),
    \label{E_approximation}
\end{equation}
where $C(\mathcal{S})$ denotes an additive constant for fixed $\mathcal{S}$.

\begin{figure}
    \centering
    \includegraphics[width=8cm]{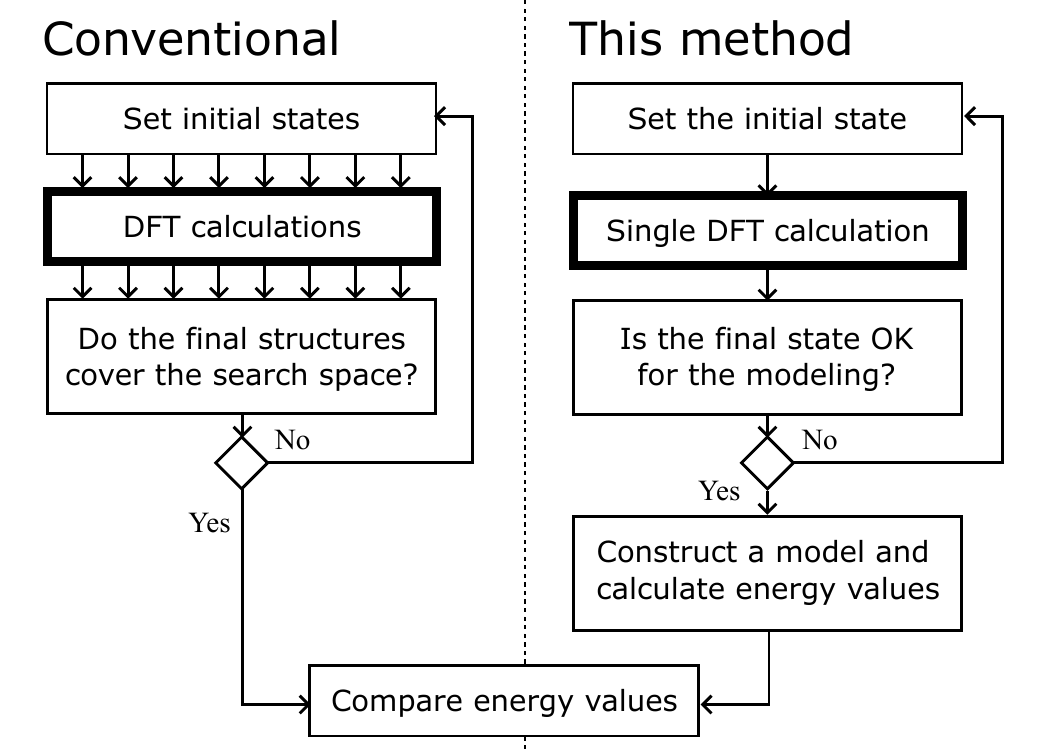}
    \caption{\label{workflow}
    Workflow for our magnetic structure search method compared with 
    the conventional scheme. 
    We discuss the occasional feedback shown in the right panel
    in Section~\ref{Results}.
    }
\end{figure}

Let $I^*$
denote the state that minimizes $E_{\mathrm DFT}(I)$.
Our assumption 
leads to the approximation with fixed $\mathcal{S}$ for $I^*$
of
\begin{equation}
    I^{*}
    \simeq
    \argmin_I
    \mathcal{E}(I,\mathcal{S})
    \label{approx_I_star}
\end{equation}
with which $I^*$ can be obtained in a single DFT calculation
with fixed $\mathcal{S}$.
Figure \ref{workflow} shows the workflow of our method compared with the conventional search
for magnetic structures. The difference is in the construction of a model that serves
$\mathcal{E}(I,\mathcal{S})$ to calculate $I^*$, which needs a single DFT calculation.
(In Section~\ref{Results}, we discuss the occasional need to
reconstruct the model).
Because DFT calculations are much more time-consuming than calculating the energy
with a Heisenberg model, our method saves computational resources.

To minimize the function with respect to $I$ in practice,
we restrict the search space by imposing sublattices on the model system.
We attribute the sublattice that is indexed by function $\ell(i)$
to the $i$th atom.
In the following results,
we use the lattice function defined as
$\ell(i)=k$ for the $i$th atom
that is equivalent 
to the $k$th atom in the unit cell
under the translation.
Let $L$ denote the number of the sublattices.
With the definition of the sublattices above,
$L$ is identical to the number of atoms in the unit cell.
Using this, we define intersublattice coupling
matrix $K_{k,l}$ as 
\begin{equation}
    K_{k,l}(\mathcal{S})
    =
    \sum_{j \in \{j|\ell(j)=l\}}
    J_{i_0,j}(\mathcal{S}).
    \label{intersublattice_J}
\end{equation}
with $i_0$ that satisfies $\ell(i_0)=k$.
$K_{k,l}$ does not depend on the choice of $i_0$ due to the translation symmetry.
 This summation can be accurately performed in the reciprocal space.
See Appendix \ref{Appendix_Reciprocal_Sum} for details.

In the search, the local magnetic states of the atoms
in a sublattice are equivalent to one another.
Let unit vector $\vec{v}_i$ denote the direction of the local moment
for the atoms in the $i$th sublattice.
We assume collinearity here, and
the state is described by an Ising model, $v_i = {\pm}1$.

The energy in the Heisenberg model can be expressed as 
\begin{equation}
    \mathcal{E}(I,\mathcal{S})
    =
    I^{\mathrm T}
    K(\mathcal{S})
    I,
    \label{Ising_energy}
\end{equation}
where
$I$ is defined as $I=(v_1, v_2, ..., v_L)^{\mathrm T}$, and
$K(\mathcal{S})$ is the matrix for which the $(i,j)$ component is
$K_{i,j}(\mathcal{S})$.

\subsection{First-principles calculations}
We perform first-principles calculations
within DFT
and the local density approximation \cite{Hohenberg64,Kohn65}
by using AkaiKKR \cite{AkaiKKR},
which is based on the Korringa--Kohn--Rostoker (KKR) \cite{Korringa47,Kohn54}
Green function method.

We assume that \ce{MnO} has a rock-salt fcc structure in 
the Fm$\bar{3}$m (\# 225) space group.
In our calculations, the lattice constant is assumed to be 8.4 Bohr.
We take the conventional 2 $\times$ 2 $\times$ 2 \ce{MnO} cell 
to accommodate several magnetic structures.
We use the 8 $\times$ 8 $\times$ 8 $k$-point mesh and reduce the number of 
$k$-points to 65 by considering the crystal symmetry in the calculation.

For \ce{SrFe12O19}, we assume that its crystal structure belongs to 
the space group P6$_3$/mmc (\#194) with lattice parameters $a=11.2$ Bohr,
and $c=43.87$ Bohr, referring to Ref.~\onlinecite{Obradors88}.
(Atoms attributed to the 4e site with an occupancy of 0.5 in the paper
are treated as the 2b site elements with an occupancy of 1
in our calculations).
We use a 6 $\times$ 6 $\times$ 2 $k$-point mesh and reduce the number of 
$k$-points to 14 by considering the crystal symmetry.

We estimate the intersite magnetic   
couplings, $J_{i,j}$, in those systems 
by using Liechtenstein's method, and the couplings are calculated
from the energy shifts by
spin rotational perturbations at the $i$th and $j$th
sites \cite{Liechtenstein87}.

\section{Results and Discussion}
\label{Results}
\subsection{\ce{MnO}}
In this subsection, we apply our method to \ce{MnO}.
MnO has four Mn atoms and four O atoms in the conventional cell.
In our application, we use the $2 \times 2 \times 2$ conventional cell,
which has 64 atoms.
A direct application of the method to 
the model yields an intersublattice coupling matrix
of $64 \times 64$ elements.
To reduce the size of the matrix,
we disregard the couplings with the O sites, and
focus on the interactions among the Mn sites.
Then, the problem becomes optimization with 
a $32 \times 32$ matrix and 32 Ising spins.

\begin{figure}
    \centering
    \includegraphics[width=8cm]{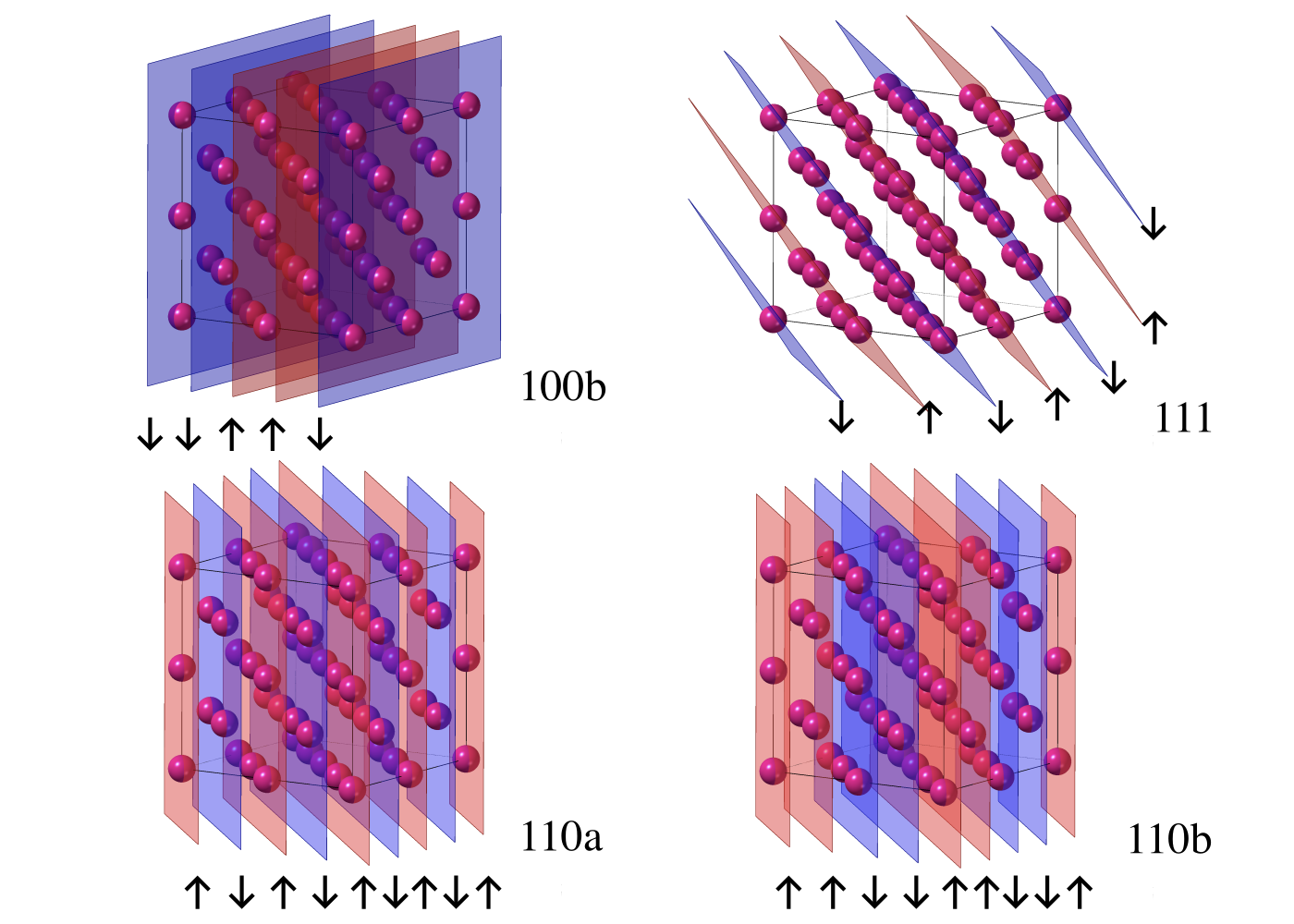}
    \caption{\label{antiferro_structure}
    Assumed magnetic structures for MnO. The local moments of the atoms on the red planes
    are directed upward, and on the blue planes they are directed downward.
    }
\end{figure}

We ease the problem further by restricting
the search space with respect to $I$
to antiferromagnetic states with 
the [100], [110], [111] directions
and the ferromagnetic state (Fig. \ref{antiferro_structure}).
The $2 \times 2 \times 2$ conventional cell can 
accommodate two [100]
antiferromagnetic structures.
Let 100a denote the structure with 
alternate layers of up and down spin,
and 100b denote an up-up-down-down
layer structure.
The cell can also accommodate two [110] antiferromagnets.
110a is the antiferromagnet with alternate layers of up and down spin,
and is identical to 100a.
110b is the antiferromagnet with an up-up-down-down layer structure.

\begin{figure}
    \centering
    \includegraphics[width=8cm]{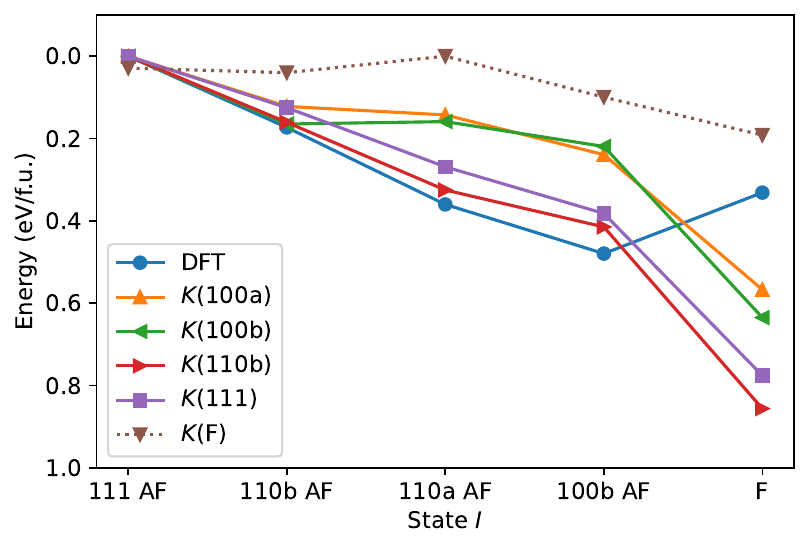}
    \caption{\label{magsearch_MnO}
    Approximated energy values as functions of the trial magnetic states, $I$  (horizontal axis).
    The values are calculated with intersublattice matrices $K(\mathcal{S})$
    from different DFT calculations with magnetic state $\mathcal{S}$.
    }
\end{figure}

Figure \ref{magsearch_MnO} shows the approximated energy
on the right-hand side of Eq.~\eqref{E_approximation}
as a function of trial state $I$.
The additive constant, $C(\mathcal{S})$, is taken so that the energy of the ground state
becomes zero.
Because the approximated energy depends on the choice of the base state,
$\mathcal{S}$, through the matrix, $K(\mathcal{S})$,
there are five different curves corresponding to the base states
(the 100a, 100b, 110b, and 111 antiferromagnetic states and the ferromagnetic
state denoted as F).
For comparison, Fig.~\ref{magsearch_MnO} also shows the energy curve from the DFT calculations.

Except for the curve calculated from the ferromagnetic base state,
all the curves have their minima at the 111 AF state,
which means that
our approximation correctly predicts the 111 AF state
as the ground state, when the base state, $\mathcal{S}$,
is antiferromagnetic.
The deviation of energy values obtained by the intersublattice matrices 
from the DFT values are adequately close so that 
the prediction of the ground state is not degraded.

Calculation with the ferromagnetic state is expected to be different
from those with the antiferromagnetic states
because the electronic structure in the ferromagnetic state
is different from the antiferromagnetic ground state of \ce{MnO}.

\begin{figure}
    \centering
    \includegraphics[width=10cm]{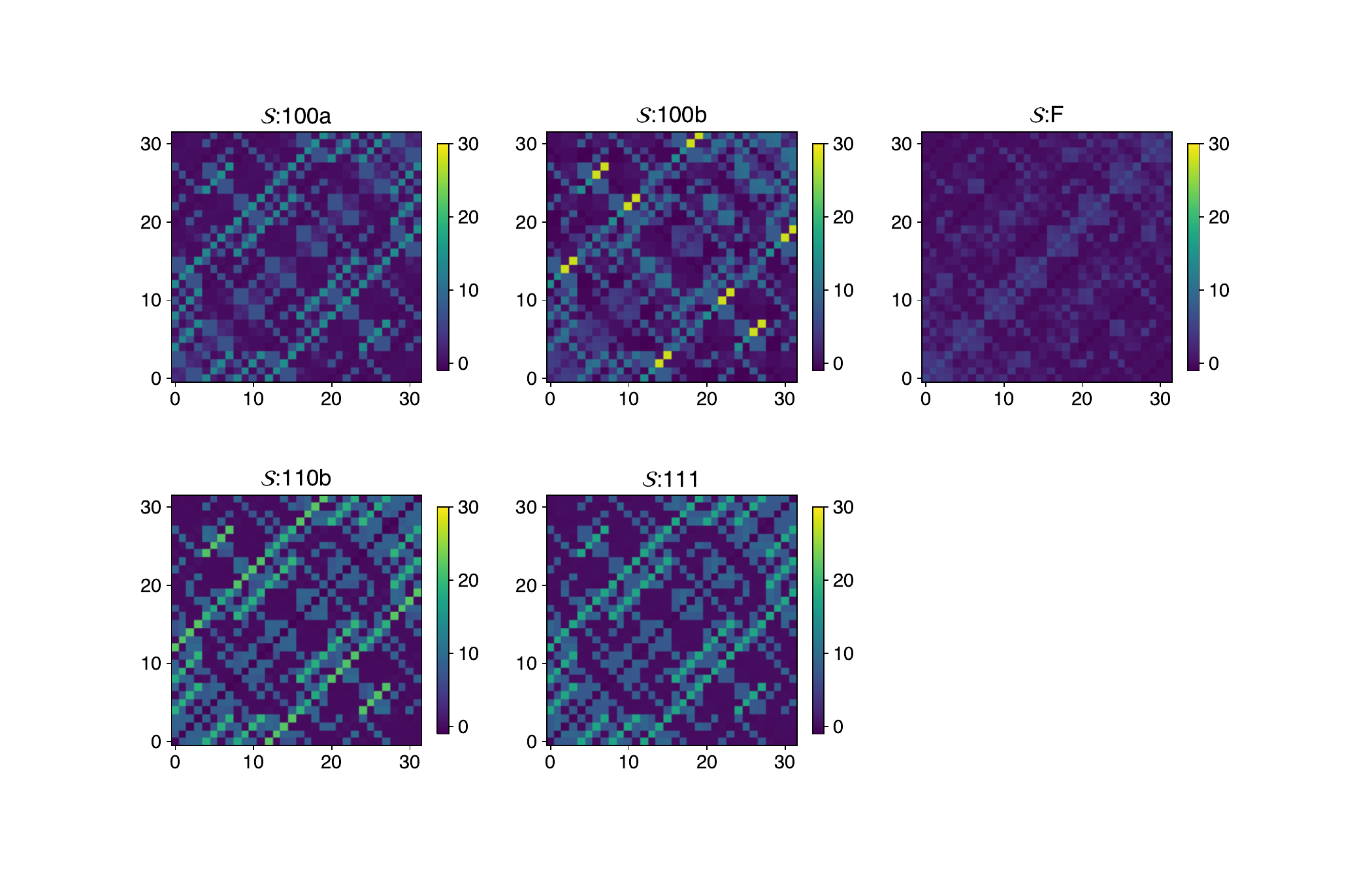}
    \caption{\label{K_matrix_MnO}
    Intersublattice matrices calculated from DFT results with 
    different magnetic states, $\mathcal{S}$.
    }
\end{figure}

We demonstrate the difference in detail from the perspective of the intersublattice
matrix, $K$.
We show 
the elements of the $K$ matrices in Fig. \ref{K_matrix_MnO}.
Most of these elements  are positive, which means most pairs
are antiferromagnetically coupled with each other.
The matrix calculated with an antiferromagnetic alignment for $\mathcal{S}$
in the DFT calculations has a similar pattern with one another.
However, those from the ferromagnetic base state have much weaker couplings.

This result probably means that the Heisenberg model
in Eq.~\eqref{eq:Heisenberg}
is adequate for describing 
the energetically lower antiferromagnetic states,
but that it is not suitable for accommodating ferromagnetic states.

\subsection{\ce{SrFe12O19}}
We use the method on \ce{SrFe12O19}
to see whether the method can be applied to
another system.
\ce{SrFe12O19} includes two formula units in the primitive cell.
We focus on the magnetic interactions among the Fe sites,
and omit the remainder elements of the intersublattice matrices.
This reduces the matrix to $24 \times 24$.
We restrict the search space for $I$
by assuming that local magnetic moments
belonging to the same Wyckoff position align in a common direction.
Because there are five Wyckoff positions for Fe in \ce{SrFe12O19}
(12k, 4f$_1$, 4f$_2$, 2b, 2a) and one of the moments can be fixed to 
the up-direction without loss of generality,
there are $2^4 = 16$ alignments to consider.

\begin{figure}
    \centering
    \includegraphics[width=9cm]{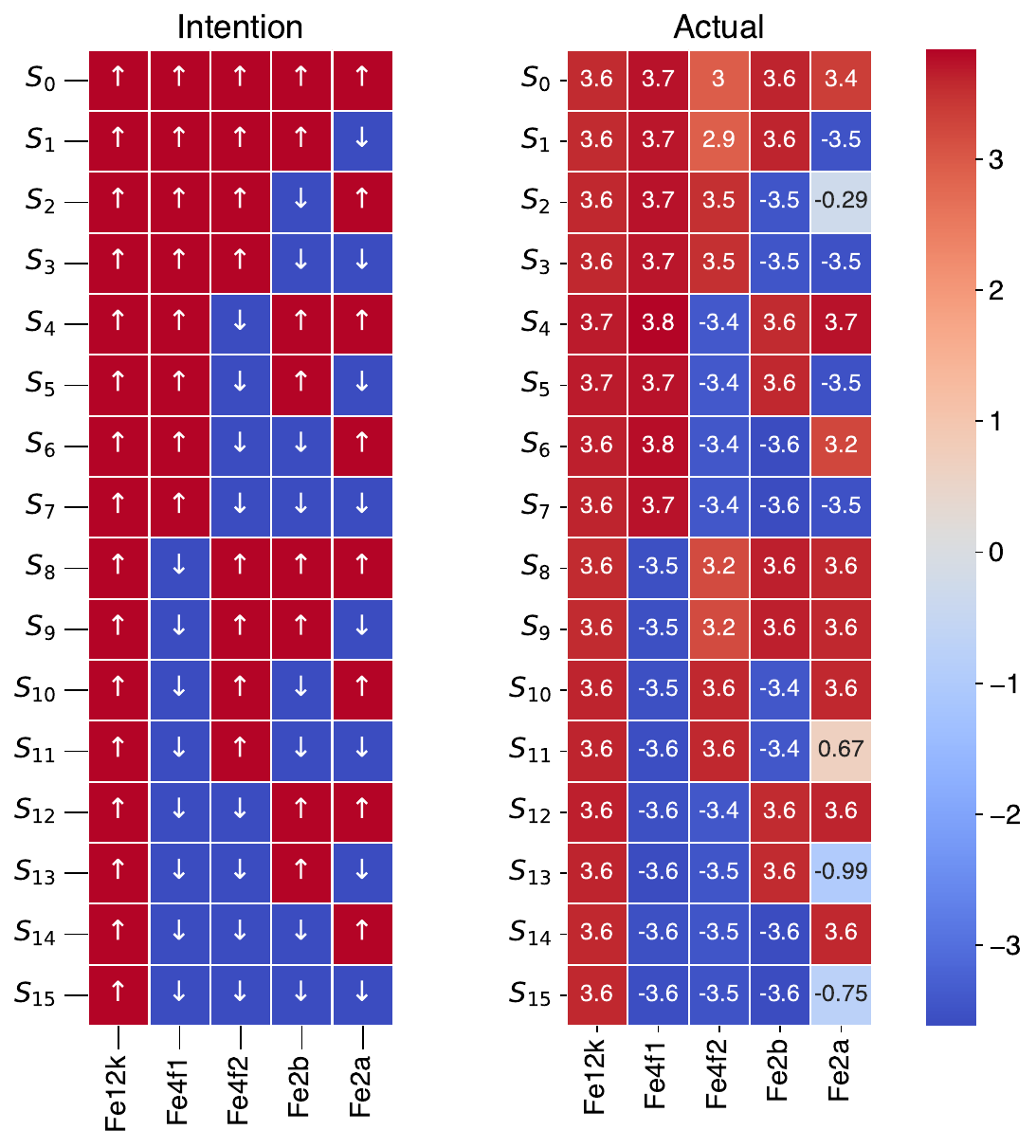}
    \caption{\label{ferrite_moments}
    Intended magnetic structures (left) and the actual local moments in $\mu_{\mathrm B}$ (right)
    obtained by DFT calculations
    in $\mathcal{S}_0,  \mathcal{S}_1, \cdots, \mathcal{S}_{15}$. 
    }
\end{figure}

We calculate intersublattice matrices with several different 
magnetic alignments for $\mathcal{S}$ as we did in \ce{MnO}.
Some of the calculations do not converge as intended
with respect to the direction of the local moments,
which is summarized in Fig.~\ref{ferrite_moments}.
The left panel shows the alignment of our intention,
and the right panel shows the final local magnetic moments
we obtained. 
For $\mathcal{S}_9$, we obtain a state identical to $\mathcal{S}_8$;
therefore, we do not consider $\mathcal{S}_9$.
We also notice that we should separate the states 
with small local magnetic moments
($\mathcal{S}_2$, $\mathcal{S}_{11}$, $\mathcal{S}_{13}$ and 
$\mathcal{S}_{15}$) in the analysis.

\begin{figure*}
    \centering
    \includegraphics[height=10cm]{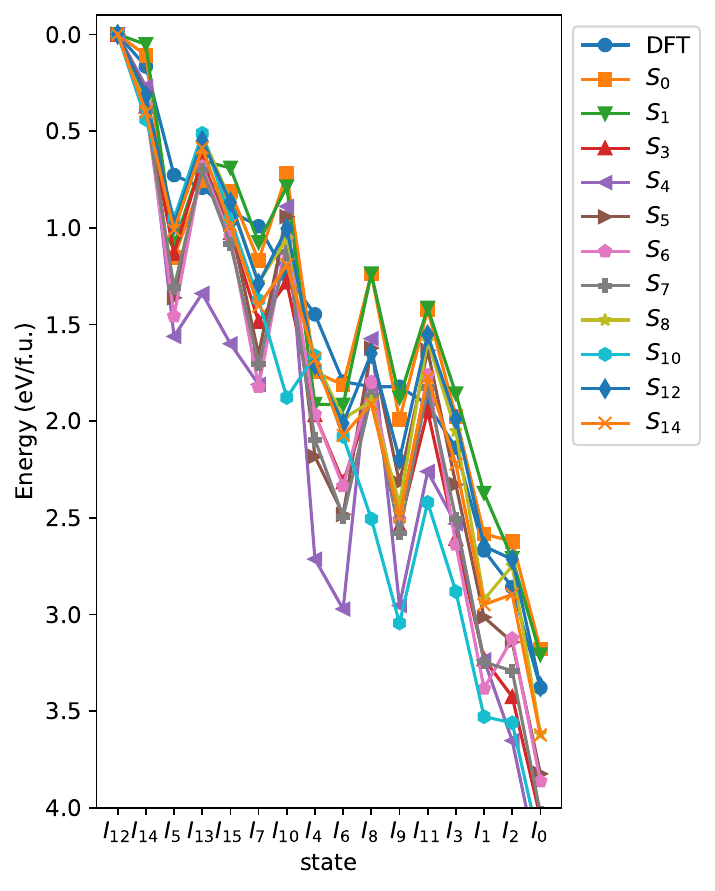}
    \includegraphics[height=10cm]{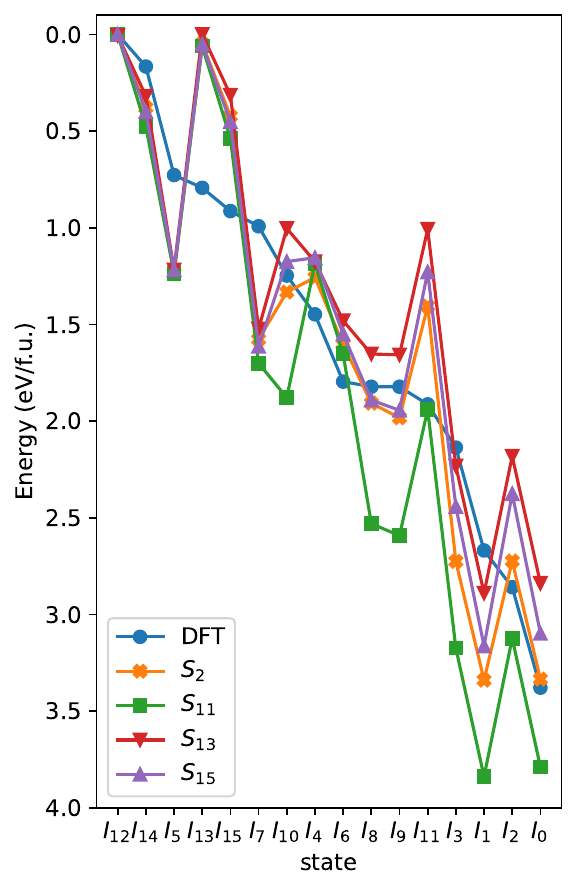}
    \caption{\label{mag_search_ferrite}
    Approximated energy values as functions of trial magnetic state $I$ (horizontal axis),
    calculated with intersublattice matrices from different DFT results.
    }
\end{figure*}

Figure \ref{mag_search_ferrite} shows the approximate energy curves
as functions of the trial state, which takes $I_1, I_2, \cdots, I_{16}$,
whose spin directions are identical to the intended directions
in $S_1, S_2, \cdots, S_{16}$ (Fig.~\ref{ferrite_moments} left).
The results from the DFT states with
intended local moments show that the approximation adequately reproduces
the energy function by DFT and predicts
the correct ground state ($I_{12}$),
as shown in the left-hand panel.

As shown in the right panel,
the predictions from the other states with unintentional local moments
are not accurate. 
This is due to underestimation of the intersublattice
interaction coming from the small 2a moments.
Consequently, the energy of the $I_{13}$ state falls 
near the energy of the ground state ($I_{12}$)
because these two states are different only in the direction
of the spin moment in 2a.

These errors can be attributed to the Heisenberg model, 
and our method seems accurate when it deals with
a system that has firm magnetic moments for transition metal sites.
Fortunately, the magnitude of the local moments can be checked
in the DFT calculation
before calculating the intersublattice 
interaction, which is a time-consuming process.
Even when we need to recalculate the system, 
our method still has a great advantage in the reduction of
the number of DFT calculations, provided that the target system
is a Heisenberg-like magnet and that
another energetic local minimum in which the local moments are 
large enough can be found easily.

\section{Conclusion}
We proposed a method for 
exploring the magnetic structure landscape with 
first-principles calculations
by using Liechtenstein's method.
Our approach allows for the efficient and accurate exploration of
complex magnetic structures with substantially reduced computational
costs.
By applying our method to MnO and hexagonal ferrites,
we demonstrated its efficiency in identifying
the ground-state magnetic structure
by using fewer first-principles calculations
than exhaustive DFT calculations.
Because our method reduces the problem 
to optimizing an Ising system, as described by 
Eq.~\eqref{Ising_energy}, we may use classical and quantum
Ising machines to solve it, which will be helpful in 
handling a large search space.
We expect that our approach is valid for Heisenberg-like magnets, including 
ferrimagnets and antiferromagnets with firm local moments,
and the method may provide a powerful tool for searching functional magnetic materials.

\section*{Acknowledgement}
We are grateful to Akihito Kinoshita, Noritsugu Sakuma, and Tetsuya Shoji
for their sound advice and fruitful discussion.
This work was supported by the ``Data Creation and Utilization-Type Material
Research and Development Project (Digital Transformation Initiative Center for
Magnetic Materials)'' project (Grant Number: JPMXP1122715503)
by the ``Program for Promoting Researches on the Supercomputer Fugaku (Computational
Research on Materials with Better Functions and Durability toward Sustainable
Development)'' project (Grant Number: JPMXP1020230325).
The calculations were conducted partly with supercomputer Fugaku provided by 
the RIKEN Center for Computational Science (Project ID: hp230205, hp240224),
and with the facilities of the Supercomputer Center at the Institute for 
Solid State Physics, The University of Tokyo.

\appendix
\section{Construction of intersublattice matrix in reciprocal space}
\label{Appendix_Reciprocal_Sum}
In this section, we show how 
the intersublattice matrix in Eq.~\eqref{intersublattice_J}
is calculated in reciprocal space with the KKR Green function method.

In Liechtenstein's formula, 
the magnetic coupling, $J_{(i,a)(j,b)}$, between the $i$th site
in the $a$th cell
and the $j$th site in the $b$th cell
is calculated
from
the scattering-path operator
and the t-matrix.
Let us denote the $i$th site
in the $a$th cell as $(i,a)$ hereafter.
The scattering-path operator, 
$T^{(i,a)(j,b)}_{L, L',\sigma}(E)$, 
of the $\sigma$-spin electrons 
from the orbital at 
the $(i,a)$ site,
whose angular momentum and 
magnetic quantum number are indexed by $L$
to that at the $(j,b)$ site with $L'$,
is a function of energy $E$.
The t-matrix, $t^i_{L,\sigma}(E)$,
of the $\sigma$-spin potential of the $i$th site
for the $L$ scattering
is also a function of $E$.
Using these equations, the coupling is formulated as
\begin{equation}
    J_{(i,a)(j,b)}
    =
    \frac{1}{4\pi}
    \sum_{L,L'}
    \Im
    \int^{\epsilon_\mathrm{F}}_{-\infty}
    d\epsilon\,
    \Delta^i_L
    T^{(i,a)(j,b)}_{L,L',\uparrow}
    \Delta^j_{L'}
    T^{(j,b)(i,a)}_{L',L,\downarrow},
    \label{Eq:Liechtenstein_Formula}
\end{equation}
where $\epsilon_\mathrm{F}$ is the Fermi energy, and
$\Delta^i_L$ the spin-rotational perturbation, defined as
$\Delta^i_L = (t^{i}_{L,\uparrow})^{-1} - (t^{i}_{L,\downarrow})^{-1}$.

With the choice of sublattice in Section \ref{method_magsearch},
$\ell((i,a))=i$,
intersublattice matrix $K$ in Eq.~\eqref{intersublattice_J}
can be written as
\begin{align}
    K_{i, j}
    =
    \sum_a
    J_{(i,0)(j,a)}
    =
    \tilde{J}_{i,j}(\vec{0})
    \label{Eq:intersublattice_matrix_app}
\end{align}
where $\tilde{J}_{i,j}$ is the Fourier transform of 
$J_{(i,0)(j,a)}$, defined as
\begin{equation}
    \tilde{J}_{i,j}(\vec{q})
    =
    \frac{1}{N_\mathrm{cell}}
    \sum_a
    J_{(i,0)(j,a)}
    e^{-i\vec{q}\cdot (\vec{R}_a-\vec{R}_0)}
\end{equation}
with the number of cells denoted by $N_\mathrm{cell}$.

Using these,  intersublattice matrix $K_{i,j}$ is expressed
in terms of the scattering path operator in the reciprocal space,
$T^{i,j}_{L,L',\uparrow}(\vec{k})$,
and the t-matrix is expressed as
\begin{equation}
    K_{i,j}
    =
    \frac{1}{4\pi}
    \Im
    \sum_{L,L',\vec{k}}
    \int^{\epsilon_\mathrm{F}}_{-\infty}
    d\epsilon\,
    \Delta^i_L
    T^{i,j}_{L,L',\uparrow}(\vec{k})
    \Delta^j_{L'}
    T^{j,i}_{L',L,\downarrow}(\vec{k}).
    \label{Eq:Liechtenstein_Formula_in_q}
\end{equation}
Because the KKR Green function method 
directly calculates
the scattering path operator,
$T^{i,j}_{L,L',\uparrow}(\vec{k})$,
this is a faster way to calculate the matrix than 
a method that detours to the real space.
Moreover, the method is free from spatial cut-offs.
The dispersion of $\tilde{J}_{i,j}(\vec{q})$,
which is related to the spin waves, can be calculated in a similar way.
We refer readers to Ref.~\cite{Fukazawa21} for details.

\bibliography{main}
\end{document}